\documentclass[aps,prl,twocolumn]{revtex4}
\usepackage{graphicx}
\usepackage{amsmath}
\usepackage{epsfig}

\begin{document}
\def\be{\begin{equation}}
\def\ee{\end{equation}}
\def\bearr{\begin{eqnarray}}
\def\eearr{\end{eqnarray}}
\def\tc{$T_c~$}
\def\tn{$T_N~$}
\def\c2{ CuO$_2~$}
\def\ruo{ RuO$_2~$}
\def\lsco{LSCO~}
\def\lco{La$_2$CuO$_4$~}
\def\lbco{La$_{2-x}Ba_x$CuO$_4$~}
\def\bi{bI-2201~}
\def\tl{Tl-2201~}
\def\hg{Hg-1201~}
\def\sro{$Sr_2 Ru O_4$~}
\def\rc{$RuSr_2Gd Cu_2 O_8$~}
\def\mgb{$MgB_2$~}
\def\pz{$p_z$~}
\def\ppi{$p\pi$~}
\def\sqo{$S(q,\omega)$~}
\def\tperp{$t_{\perp}$~}
\def\ndc{Nd$_{2-x}$Ce$_x$CuO$_4$~}
\def\nd{Nd$_{2}$CuO$_4$~}
\def\prc{Pr$_{2-x}$Ce$_x$CuO$_4$~}
\def\pr{Pr$_{2}$CuO$_4$~}
\def\lar{La$_{2-x}^{3+}$R$^{3+}_x$CuO$_4$~}
\def\lbco{${\rm La_{2-x}Ba_xCuO_2}$~}
\def\lczno{${\rm La_{2}Cu_{1-x}Zn_xO_2}$~}
\def\cef{Ce$^{4+}$~}
\def\ce3{Ce$^{3+}$~}
\def\cut{Cu$^{2+}$~}
\def\half{$\frac{1}{2}$~}
\def\cu1{Cu$^{1+}$~}
\def\cu2{Cu$^{2+}$~}
\def\Cef{Ce 4$f_{z(x^2-y^2)}$~}
\def\dx2{Cu 3$d_{x^2-y^2}$~}
\def\tj{$\rm{t-J}$~}
\def\hlf{$\frac{1}{2}$~}

\title{Superconductivity in optimally doped Cuprates:\\
BZA Program works well \& Superexchange is the Glue}

\author{G. Baskaran}
\affiliation{Institute of Mathematical Sciences\\ Chennai 600 113, India}
\date{\today}

\begin{abstract}
Resonating valence bond states in a doped Mott insulator was proposed to explain superconductivity in cuprates in January 1987 by Anderson. A challenging task then was proving existence of this unconventional mechanism and a wealth of possibilities, with a rigor acceptable in standard condensed matter physics, in a microscopic theory and develop suitable many body techniques. Shortly, a paper by Anderson, Zou and us (BZA) undertook this task and initiated a program. Three key papers that followed, shortly, essentially completed the program, as far as superconductivity is concerned: i) a gauge theory approach by Anderson and us, that went beyond mean field theory ii) Kotliar's d-wave solution in BZA theory  iii) improvement of a renormalized Hamiltonian in BZA theory, using a Gutzwiller approximation by Zhang, Gros, Rice and Shiba. In this article I shall focus on the merits of BZA and gauge theory papers. They turned out to be a  foundation for subsequent developments dealing with more aspects that were unconventional - d-wave order parameter with nodal Bogoliubov quasi particles, Affleck-Marston's $\pi$-flux condensed spin liquid phase, unconventional spin-1 collective mode at $(\pi,\pi)$, and other fascinating developments. Kivelson, Rokhsar and Sethna's idea of holons and their bose condensation found expression in the slave boson formalism and lead to results similar to BZA program.

{\em At optimal doping, correlated electrons acquire a `reasonable' fermi sea, at the same time retain enough superexchange inherited from a Mott insulator parentage, ending in a BCS like situation with superexchange as a glue !} Not surprisingly, mean field theory works well at optimal doping, even quantitatively. Further, t-J model is an excellent minimal model around optimal doping, where RVB superconductivity is also at its best. 
\end{abstract}

\maketitle

\begin{center}
\textbf{\large Introduction}
\end{center}

Discovery of high \tc superconductivity in cuprates by Bednorz and M\"{u}ller \cite{bednorz} in 1986 is a remarkable event. It was a breakthrough and a major turning point in the history of superconductivity and strongly correlated electron systems. The field of quantum condensed matter physics and the community even got reorganized.  About a month after Tanaka and collaborators confirmed\cite{tanaka} and brought to light Bednorz-M\"{u}ller's path breaking discovery, one of Anderson (PWA) \cite{pwascience} proposed a theory for high-$T_c$ superconductivity. We followed it up with several papers\cite{BZA,gauge1,pwa1,za,wha,pwabook}, focusing on superconductivity, within a span of one year. Several groups joined hands resulting in a flood of activities.

Anderson's paper, to be referred to as PWA\cite{pwascience}, which suggested the resonating valence bond (RVB) mechanism of superconductivity, was very appealing and at the same time very unconventional. A challenge then was, as a first step, i) theoretically establishing RVB mechanism of superconductivity, in the t-J model and ii) develop a suitable many body theory that will be useful to calculate a variety of physical quantities and also confirm several of the predictions of PWA. This task was executed in two papers, in quick succession, by Anderson Zou and us (BZA)\cite{BZA} and Anderson and us (BA)\cite{gauge1}. During this period, except for a large value of superconducting \tc, not much was known experimentally, including fundamental properties such as, number of bands crossing the fermi level, symmetry of superconducting order parameter and survival of superexchange interaction in the doped Mott insulator. It is in this background, these two papers provided a proof, acceptable in standard condensed matter physics, of the unconventional phenomena of RVB superconductivity in a doped Mott insulator, and a theoretical frame work, a program, that became foundational for further quantitative developments. This was the beginning of the BZA program. 

The richness of Anderson's suggestion was that Superconductivity in cuprates has turned out to be unconventional in more than one way: i) a new electronic mechanism, in an unexpected place called a doped Mott insulator and associated large \tc, ii) preformed spin pairing, later called
`spin gap' phenomenon, an unusual precursor to superconductivity, over a wide temperature region above \tc, iii) unconventional d-wave order parameter with nodal quasi particles, iv) unconventional spin-1 collective mode at $(\pi,\pi)$ and v) an unusual competition from other types of charge and spin orderings etc.

Superconductivity in cuprate family is a robust phenomenon at optimal doping. 
It has overcome disorder, charge and spin order tendencies and lattice instabilities. At optimal doping it is present in all cuprates containing CuO$_2$ layers. A large condensation energy is evident in the way superconducting \tc, at optimal doping, jumped from the range of 30 to 90 to 120 and then 160 Kelvin, in new members of the cuprate family. A record \tc $\sim 163 K$ is being held by a Tl based cuprate under a large external pressure. As RVB theory was based on spins and their exchange interactions, it was able to account for the large transition temperature and large condensation energies in a natural fashion, compared to attempts based on phonon and other mechanisms. A strength of RVB theory from the beginning was its sound phenomenological basis, from where a flow of new concepts was natural. Mathematical difficulties that followed were in the nature of strongly correlated electrons in a tight binding band; a suitable many body theory did not exist. Interestingly, these formidable mathematical difficulties were also overcome, very efficiently, in the theoretical developments that quickly followed. 

There have been efforts, before Bednorz-M\"{u}ller's discovery, in discussing possibility of superconductivity in models containing repulsive interactions, to understand superconductivity in heavy fermions and organics.  Historically, Hirsch\cite{hirsch1} was the first to suggest an extended-s pairing in a repulsive Hubbard model. In a subsequent paper Scalapino, Loh and Hirsch\cite{hirsch2} interpreted the same superconductivity as spin fluctuation mediated pairing. Other authors have used the idea of d-wave pairing mediated by spin fluctuations in nearly antiferromagnetic metals, such as heavy fermions\cite{varma} and Bechgard salts\cite{emery}. The idea of pairing due to spin fluctuations continue to be pursued for high \tc cuprates by several groups\cite{pines,moriya}. One of the aims of the present paper is to bring out, as clearly as possible, superexchange rather than exchange of spin fluctuations is a natural, physically correct and mathematically straight forward method to describe the glue for cuprates. \textbf{Mott insulator is the template and superexchange is the glue.}

There has been a variety of efforts with varying success, in studying directly t-J and Hubbard models in 2D, for superconductivity, along the RVB route: Kivelson, Rokhsar and Sethna's idea\cite{KRS} of soliton doping and bose condensation of holons, detailed slave boson analysis\cite{za,kotliar,kotliarLu},
detailed work\cite{fukuyama} that sharpened the BZA phase diagram, an improved renormalized Hamilonian analysis\cite{ZGRS}, gauge theory 
approaches\cite{piFlux,matsui,wiegman,gauge2,ioffe,wwz,plee,sachdev,tesanovic,senthil,Ng,wenBook,pleeReview}, variational monte-carlo\cite{paramekanti}, quantum monte carlo\cite{sorella}, k-space\cite{jarrel} and real space\cite{dmftKotliar} cluster DMFT methods, diagrammatics\cite{tremblay,maierScalapino} powerful renormalization group studies of Hubbard model in 2D\cite{riceRG,metzner}, exact diagonalization\cite{white,exactDiag,prelovcek}, series expansion\cite{putikka,ogata2} 
and some analytical\cite{analytical}methods
have been employed.

In the following sections, we elaborate a view that the basic and important problem of establishing an electronic mechanism of superconductivity by a many body theory and key physics of Mott insulators, during 1987, was solved in the very first phase. Anderson has expressed this view \cite{vanila,pwaSpinLock} in a recent article. The present article echoes similar views from a slightly different perspective giving some details. It has bit of history, as we will be completing 20 years since Bednorz-M\"{u}ller's path breaking discovery. 

There is a recent review article by Lee, Nagaosa and Wen\cite{pleeReview}, which discusses physics of high \tc superconductivity in doped Mott insulators. It touches many of the theoretical developments. Our focuss is on the BZA program, which has been so far very successful and has the potential to be used extensively for further quantitative progress. We also give some new insights and discussions.  

In the concluding section we contrast and distinguish, theory of superconductivity in elemental metals from cuprates and other potential RVB superconductors. Even the type of questions raised and the way one addresses issues are different. The scientific efforts put in unraveling the mystery of the complex cuprate system does not have many parallel in condensed matter physics. At the end, it is fair to say we do understand a lot, to be able to say where superconductivity comes from and what is the mechanism. There is more to be understood, of course. Such a realization has two effects: i) some what loose statement one hears occasionally, `even now we do not understand high \tc superconductivity and the mechanism remains still unclear', looses its validity and  ii) it gives one confidence and suggests that RVB theory is well and the BZA program is ready to answer new and old questions from experiments. Some encouraging recent examples from theory are: i) variational Monte Carlo analysis of Gutzwiller projected RVB wave functions by Nandini, Paramekanti and Randeria\cite{paramekanti}, and finding a good agreement with results of the BZA program as well as some experimental results ii) detailed calculation of electronic structure properties and excited state properties by Gros-Muthukumar group\cite{muthuGros}, Ogata's group\cite{ogatta} Zhang's group\cite{zhangGroup} and others and iii) Anderson's very recent attempt to describe superconductivity and spin gap phenomena in an unified fashion using a notion of spin-charge locking and two types of Anderson-Nambu spinors.

\begin{figure}[h]
\includegraphics[width=9cm]{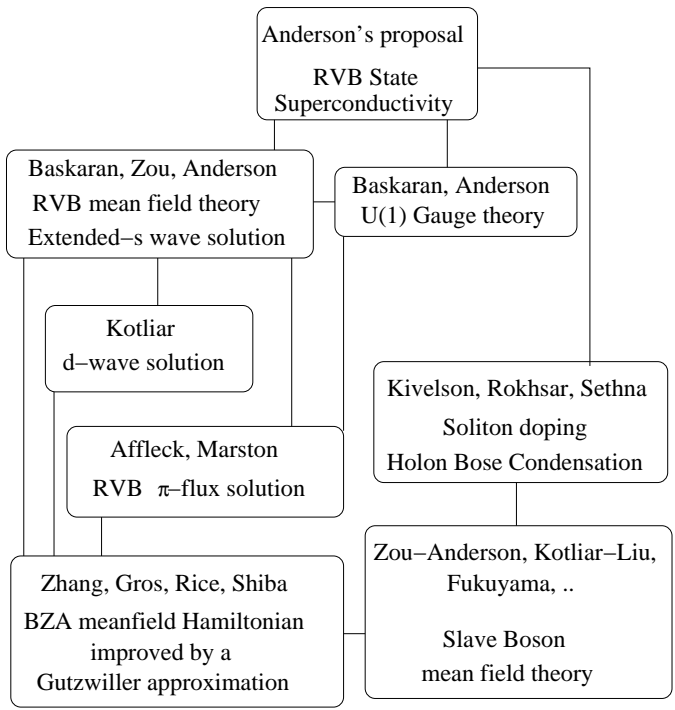}
\caption{Evolution of BZA program beginning with Anderson's proposal.
Kivelson et al.'s bose condensation of holon is an independent development
that also contributed to BZA program through slave boson approach.}
\end{figure}

\begin{center}
\textbf{\large{Brief Introduction to a Trio}}
\end{center}

RVB proposal\cite{pwascience} in January 1987 created a spontaneous involvement of theorists and experimentalists from all over the world, from widely different background. The idea flourished instantly. Many key developments took place during 1987-88. For example, Kivelson, Rokhsar and Sethna's idea\cite{KRS} of soliton doping and bose condensation of holons, spin-charge decoupling, Kotliar's d-wave solution\cite{kotliar}, Zou-Anderson's slave boson formalism\cite{za} adapted to t-J model, Affleck-Marston's phase\cite{piFlux}, a condensate of $\pi$-flux of RVB gauge field, enlargement of U(1) to SU(2) RVB gauge theory\cite{su2}, Zhang-Rice singlet construction\cite{zhangRice}, an improvement of the renormalized Hamiltonian used in BZA theory\cite{ZGRS}, sharpening of BZA phase diagram by detailed slave boson studies\cite{kotliarLu,fukuyama}, statistics transmutation\cite{polyakov}, Laughlin's idea\cite{semionSC} of semion superconductivity (condensation of holons carrying an RVB gauge field flux of $\frac{\pi}{2}$), other PT violating chiral RVB states\cite{kalmayer} Chern-Simons gauge fields, computation of physical quantities using RVB gauge field theory, anomalous normal state, failure of fermi liquid theory, electron confinement, interlayer pair tunneling and more. Some of them that are directly related to one layer superconductivity are shown in figure 1.

\begin{table}
\begin{tabular}{|p{9mm}|p{13mm}|p{16mm}|p{17mm}|p{13mm}|p{10mm}|}
\hline
Theory& Reference State & Mechanism & Methods & Exci- tations & Pheno- meno- logical Theory \\
\hline\hline
BCS    & Fermi Liquid & Exchange of Bose quanta & BCS, Nambu-Gorkov,
Eliashberg,
BdG, etc. & Bogol- iubov QP & GL\\
\hline
RVB & Mott Insulator (Spin-Liquid State) & Super Exchange is the Glue
& RVB-MFT, RVB-Gauge Theory& Neutral Fermions? & RVB-GL \\
\hline
\end{tabular}
\caption{Comparison between conventional BCS theory
and 1987 RVB theory. BCS = Bardeen-Cooper-Schrieffer theory, RVB = Resonating
Valence Bond theory, QP = quasi particles, GL = Ginzburg-Landau theory, BdG =
Bogoliubov de Gennes Equations, MFT = mean-field-theory.}
\end{table} 
 
These intense activities revolved around PWA, BZA and BA papers. We call these `trio', as these three papers have a close link and continuity. The RVB superconducting state and related theoretical developments in the trio in 1987 is similar to a BCS type of theory, but based on electron correlation mechanism, with its own novel features and notions and some formidable theoretical problems. This is shown in Table 1. After phonon pairing mechanism and BCS theory, RVB theory is the most significant development in the field of superconductivity, involving an entirely different mechanism based on electron correlations and more importantly compelled by a rich phenomenology. Even though heavy fermion superconductors and organic superconductors existed around 1987, with an electronic mechanism at work, it did not excite the condensed matter community as much as cuprates did.

\begin{flushleft}
\textbf{1. Anderson's Proposal}
\end{flushleft}

A qualitative and quantitative understanding of high Tc superconductivity
in cuprates involved, first identifying the predominant mechanism of superconductivity. This, in turn, involved three major steps: i) abstracting key notions and introducing a new paradigm, ii) identifying the right low energy effective Hamiltonian and iii) developing suitable theoretical methods and finding approximate solutions. Finding a new paradigm, abstracting new notions and a model for cuprate superconductivity, using known phenomenology of \lco and other magnetic oxides is a remarkable chapter in condensed matter physics. As PWA has expressed
in an article\cite{magnetician} entitled `Magnetician's edge', a detailed and in depth knowledge
of quantum magnetism (particularly in oxides) was essential. The phenomena of cuprate superconductivity turned out to be a meeting ground of quantum magnetism and superconductivity.  The spirit of the approach to this complex quantum condensed matter problem is well summarized in `Central dogma in high \tc superconductivity', a chapter in PWA's book on high \tc superconductivity\cite{pwabook}. 

PWA\cite{pwascience} identified the parent compound \lco as a spin-\hlf Mott insulator, having one electron in a non-degenerate orbital per copper atom.
This as well as the doped \lbco is described by a single band large U Hubbard 
model in 2D at and away from half filling:
\be
H = - \sum_{\langle ij\rangle}t^{}_{ij}  c^\dagger_{i\sigma} c^{}_{j\sigma} + h.c. +
U \sum_i n^{}_{i\uparrow}n^{}_{i\downarrow}
\ee 
Here the site index refers to a Wannier orbital. It is a symmetry adapted
hybrid of  \dx2 and oxygen 2p orbitals, that retains d$_{x^2-y^2}$ symmetry.
At half filling and when U $>>$ t, the ground state is a Mott insulator. It has
a finite Mott Hubbard gap for charge carrying excitations. Various estimates of
U and t and also next nearest neighbor hopping's exist now: U $\sim$ 5 eV,
t $\sim$ 0.25 eV. The ground state of the Hubbar model at half filling is well approximated by the ground state for hopping matrix element t = 0 :
\be
|\sigma_1,\sigma_2, ..., \sigma_N\rangle \sim \prod_{i=1 to N} c^{\dagger}_{i\sigma_i} |0\rangle
\ee
In these states, every site is singly occupied and has a dangling spin. Consequently, total spin degeneracy of this manifold is $2^N$. The extensive spin entropy of the above states are removed by superexchange, a second order hopping processes, involving two neighboring sites at a time. By a second order perturbation procedure we can derive an effective Hamiltonian that lifts the 2$^N$ fold spin degeneracy. For a given pair of neighboring sites, the
ground states for hopping, t = 0,  are:

\be
|\uparrow, \downarrow\rangle,~~|\downarrow, \uparrow\rangle,~~|\uparrow, \uparrow\rangle,~~|\downarrow,\downarrow\rangle,
\ee

These 2$^2$ neutral spin configurations are three bond triplets and a bond singlet state of two spins of neighboring sites. When hopping t is introduced perturbatively, there is a virtual transition or mixing of the above states with the excited `ionic spin singlet' intermediate configurations:
\be
|\uparrow \downarrow, 0\rangle {\rm ~and~} |0, \uparrow \downarrow\rangle ,
\ee
resulting in a bond singlet ground state and bond triplet excited state. This is 
represented by the following effective Hamiltonian for the spin dynamics of
the large U repulsive Hubbard model:
\be
H{\rm(half~filling)} \rightarrow 
H_s = J\sum_{\langle ij\rangle} ({\bf S}_i \cdot {\bf S}_j - \frac{1}{4})
\ee
where $ J = \frac{4t^2}{U}$. 

PWA suggested that superexchange survives in the doped Mott insulators, up to some reasonable dopings. Electrons delocalize, but continue to respect the double occupancy constraint. The resulting model for non-half filled case is t-J model that contains in addition to the superexchange term H$_s$ also the kinetic energy term H$_t$:
\bearr
H_{tJ} &=& H_t + H_s \\
&=& - \sum_{\langle ij\rangle}t^{}_{ij}  c^\dagger_{i\sigma} c^{}_{j\sigma} + h.c. +
J\sum_{\langle ij\rangle} ({\bf S}_i \cdot {\bf S}_j - \frac{1}{4}n_in_j) \nonumber
\eearr
with a double occupancy constraint, n$_{i\uparrow}$ + n$_{i\downarrow} \neq$ 2 at every site. Cuprates have a large superexchange J $\sim$ 0.15 ev, one of the largest among spin-\hlf Mott insulators.

PWA further suggested, in view of strong quantum fluctuations arising from spin-\hlf character and 2 dimensionality, that ground state of this 2D Heisenberg model is  a quantum spin liquid, with a possible psuedo fermi surface for certain neutral fermion excitations. The magnetic susceptibility data of Ganguly and Rao\cite{ganguly}, for insulating \lco, which did not exhibit any phase antiferromagnetic phase transition feature, also seemed to support PWA's earlier notion of spin liquids\cite{pwa73} in 2D spin-\hlf Heisenberg antiferromagnets. 

Singlet correlations in this quantum spin liquid was suggested as the neutral singlets or preformed pairs that are waiting to superconduct, given an opportunity. On doping, a fraction x of neutral resonating singlets get charged resulting in superconductivity. Here x is the doping fraction. The RVB mechanism was expressed succinctly in the form of a Gutzwiller projected (double occupancy removed) BCS type wave function 
\bearr
|{\rm RVB}; \phi\rangle &=& P_G \prod_k
(u_k + v_k c^{\dagger}_{k\uparrow}c^{\dagger}_{-k\downarrow})|0\rangle
\nonumber \\
&\equiv& P_G [\sum_{ij} \phi (ij)
b^{\dagger}_{ij} ]^{\frac{N(1-x)}{2}}|0\rangle
\eearr
that nicely interpolates the spin liquid ground state of the Mott insulator
(x = 0) and the superconducting doped Mott insulator (x $\neq$ 0). Here N is the number of sites and N(1-x) is the number of electrons. Gutzwiller projection is defined as,
\be
P_G \equiv \prod_i (1-n_{i\uparrow}n_{i\downarrow})
\ee

There was another important suggestion in this paper: two electrons in a given cooper pair will avoid double occupancy and cooper pair function $\phi(ij)=0$ 
for i = j. This automatically allowed extended-s and higher angular momentum symmetry such as d-wave. 

This paper was very special. Superconductivity emerged from a Mott insulator (non-fermi liquid) background. Pairing was not in k-space: superexchange, an intrinsically real space quantum chemical binding phenomenon lead to zero momentum condensation of cooper pairs. This paper\cite{pwascience} has become a classic in quantum condensed matter physics, almost a poem that opened a new door and one that reveals new shades of meaning each time one reads it.

\begin{flushleft}
\textbf{2. BZA theory}
\end{flushleft}
BZA followed heels and provided a physically motivated approximation method for the strongly correlated electron problem at hand. It ended up being a beginning of a program. As mentioned earlier, PWA suggested a Gutzwiller projected variational wave functions parametrized by a pair function $\phi(ij)$. This theory undertook this variational analysis. \textit{This is similar to a BCS-Hartree Fock type analysis, but in a restricted Hilbert space containing no double occupancy.} That is, one would like to minimize the energy expectation value (or free energy) with respect to the pair function $\phi(ij)$:
\be
E[\phi] = \langle {{\rm RVB};\phi|P_G (H_t + H_s) P_G |{\rm RVB}}; \phi \rangle
\ee
Presence of Gutzwiller projector P$_G$ makes computation formidable. This theory introduced a physically motivated approximation. The approximation amounts to treating the Gutzwiller projection in a mean field fashion and approximate the above expression by 
\be
E[\phi] \approx \langle {{\rm RVB};\phi|(xH_t + H_s)|{\rm RVB}}; \phi \rangle
\ee
That is, the complicated Gutzwiller projection was approximated by replacing the hopping parameter t by a renormalized parameter xt, since x is the probability that an electron can find a neighboring site empty to which it can hop.  In other words we have a renormalized Hamiltonian
\be
{\tilde{H}}_{tJ} = xH_t + H_s
\ee
defined in the full Hilbert space, also containing double occupancies. The rest is very similar to standard BCS theory. Interestingly this paper conjectured that this renormalization prescription should work well beyond about 5$\%$ doping, about which we will discuss later.

Within the above mentioned approximation this theory found i) a spin liquid ground state, neutral fermion excitations with a pseudo fermi surface for the Mott insulator and ii) a superconducting ground state with extended-s symmetry for doped Mott insulator.

A key point in BZA paper is a liberation from the Pauli spin operators, that is traditionally used for analysis of quantum spin systems and go to the constituent electron variables, even for the Mott insulator, by rewriting the Heisenberg part of the Hamiltonian as
\be
 H_s = J\sum_{\langle ij\rangle} ({\bf S}_i \cdot {\bf S}_j - \frac{1}{4}) = -J
 \sum_{\langle ij \rangle} b^{\dagger}_{ij}b^{}_{ij}
\ee
using the relation, ${\bf S}_i \equiv \sum_{\alpha, \beta} c^{\dagger}_{i\alpha}
\tau_{\alpha\beta} c^{}_{i\beta}$, where $\tau$ is the Pauli spin operator. Further $b^{\dagger}_{ij} \equiv \frac{1}{\sqrt2} (c^{\dagger}_{i\uparrow}c^{\dagger}_{j\downarrow}-
c^{\dagger}_{i\downarrow}c^{\dagger}_{j\uparrow})$ is the bond singlet or
(in the present case) neutral cooper pair operator.
In the electron representation the Heisenberg Hamiltonian has a simple
meaning. The spin-spin coupling encourages bond singlets, because it is
minus of the bond singlet number operator $b^{\dagger}_{ij}b^{}_{ij}$.
The non-trivial character of the lattice problem arises from the fact
that the bond singlet number operators do not commute, if they share one 
common site. We \cite{vbam} showed recently a very useful commutation relation, 
\be[b^{\dagger}_{ij}b^{}_{ij},b^{\dagger}_{jk}b^{}_{jk}] = {\bf S}_i \cdot ({\bf S}_j \times{\bf S}_k)\ee. It has a deep meaning that singlet resonance or delocalization involves an unavoidable spin chirality fluctuation.

In k-space the cooper pair scattering term arising from superexchange has the following form:
\be
H_{\rm pair} = -J\sum_{k,k'} \gamma({\bf k - k'}) c^{\dagger}_{-k'\downarrow}c^{\dagger}_{k'\uparrow}c^{}_{k\uparrow}c^{}_{-k\downarrow}
\ee
with the pair potential having the form, $\gamma({\bf k - k'}) \sim [\cos(k_x-k_x') + \cos(k_y-k_y')]$. It should be noted that the pair potential, while it is attractive for small momentum transfer $({\bf k - k'}) \sim 0 $ , changes sign and becomes repulsive for large momentum transfer $\sim (\pi, \pi)$, manifestly suggesting a d$_{x^2-y^2}$-wave rather than extended-s wave as a low energy mean field solution. For a reason that will be elaborated later, we were extremely satisfied with extended-s wave mean field solution.

It then employed a Bogoliubov-Hartree-Fock factorization and identified nearest neighbor self consistent parameters:
\bearr
\Delta &\equiv& \sum_k 
(\cos k_x + \cos k_y)\langle c^{\dagger}_{k\uparrow}c^{\dagger}_{-k\downarrow}\rangle
{\rm ~~~~and} \nonumber \\
p &\equiv& 
\sum_{k\sigma} 
(\cos k_x + \cos k_y)\langle c^{\dagger}_{k\sigma}c^{}_{k\sigma}\rangle ~~~~~~~~~~~~
\eearr
The first one $\Delta$ is the usual anomalous superconducting amplitude. The second one p is somewhat unconventional, it is a kinetic energy or hopping term, a `Hartree-Fock Vector Potential', generated by superexchange process. This unusual Hartree-Fock factorization term introduced in this paper played crucial role in later developments, such as gauge theory\cite{gauge1} and Affleck-Marston's flux phase\cite{piFlux}.  

\begin{flushleft}
\textbf{2.1 Mott insulator}
\end{flushleft}
Let us first consider the case of zero doping, x = 0, the Mott insulator.
The simplest self consistent solution was found to be $\Delta = 1$ and p = 0.
This resulted in a quasi particle Hamiltonian for neutral fermions, $\alpha$'s :
\be
H_{\rm mF} \sim J \sum_{k\alpha} 
|\cos k_x + \cos k_y|~~ \alpha^{\dagger}_{k\sigma} \alpha^{}_{k\sigma}
\ee
The pseudo fermi surface for the neutral fermions is given by the expression
$ |\cos k_x + \cos k_y| = 0$. Further, the anomalous pairing leads to a remarkable result for ground state occupancy
\be
  n_{k\sigma} \equiv \langle c^{\dagger}_{k\sigma} c^{}_{k\sigma}\rangle = 1
\ee
Even though neutral fermion excitations have a pseudo fermi surface,
there is no momentum space discontinuity for the constituent electrons. In this sense this spin liquid ground state of the Mott insulator is far removed from
any standard fermi liquid state. The ground state suggested by this RVB mean 
field theory is the Gutzwiller projected spin liquid state:
\be
|{\rm 2D~RVB}\rangle = P_G \prod_k
(u_k + v_k c^{\dagger}_{k\uparrow}c^{\dagger}_{-k\downarrow})|0\rangle
\ee
with $\frac{v_k}{u_k} = \pm 1$ inside and outside the pseudo fermi surface
respectively. Emergence of neutral fermions with a pseudo fermi surface, in a many body theory for quantum spins, was a great excitement at that time. It was radically different from bosonic spin wave excitations in ordered antiferromagnets.

This paper also pointed out that in the Mott insulator, `neutral fermion' quasi particles are meaningful only when they are created as `particle-hole' pairs $P_G \alpha^\dagger_{q\sigma}\alpha^{}_{q'\sigma'}\prod_k
(u_k + v_k c^{\dagger}_{k\uparrow}c^{\dagger}_{-k\downarrow})|0\rangle $, with q and q' definited on opposite side of the fermi surface. When they are on the same side of the fermi surface we end up creating charged fermions, which are not part of the low energy excitation spectrum of the Mott insulator. This implied that Brilluoiun Zone of neutral fermions are only half of the full BZ, very much like the BZ of spinons in the case of 1D Heisenberg spin-\hlf antiferromagnet.

\begin{flushleft}
\textbf{2.2 RVB superconductor}
\end{flushleft}

The mean field analysis was then performed for the doped Mott insulator, x $\neq$ 0. In addition to the superexchange term it had the renormalized kinetic energy term
$x H_t$ (equation 11). The problem at this level becomes very much like a standard BCS analysis, with a renormalized band width and nearest neighbor cooper 
pairing of strength J. Thus superexchange becomes the glue.

In addition to the anomalous average $\Delta \neq 0$ (unlike the case of Mott insulator) it also found p $\neq$ 0. It is a renormalization of kinetic energy from superexchange term. The overall superconducting solution corresponds to a spin singlet superconducting state with extended-S symmetry.

The mean field solution suggested superconductivity (ODLRO) for any filling. However, at half filling, ODLRO is destroyed by a complete suppression of 
low energy charge fluctuations, implemented by Gutzwiller projection.
That is, at half filling the mean field superconducting state turns into a 
neutral spin liquid state upon elimination of charge fluctuations.
The mean field solution suggested superconductivity (ODLRO) for any filling. However, at half filling, ODLRO is destroyed by a complete suppression of 
low energy charge fluctuations, implemented by Gutzwiller projection.
That is, at half filling the mean field superconducting state turns into a 
neutral spin liquid state upon elimination of charge fluctuations.

It was pointed out that the mean field transition temperature obtained by this
analysis, which has the generic BCS form, 
\be
k_BT_c^* \sim J
\ee
a fraction of the superexchange J, is the cross over temperature at which `preformed spin singlet pairs' are beginning to appear in the system. In current terminology this temperature scale is the `spin gap' scale, below which spins are progressively getting paired in a cooperative fashion. These singlet pairs start transporting charge and compete with single electron transport. Thus the mean field temperature provides some kind of umbrella below which the charged valence bonds can undergo BEC type of condensation on their own, making use of their light mass and small density. For small doping the actual superconducting transition temperature will be bounded by the mean field \tc. 

As doping x increases, the effect of superexchange is decreasing and at the same
time a fermi sea is being built, because of increase in electron delocalization. Thus we have some kind of fermi sea with superexchange as the glue. So the \tc predicted by mean field theory will become close to the superconducting \tc. This is shown in figure 2, which reproduces the BZA mean field phase diagram. Beyond the dashed line, mean field results for superconducting \tc starts making sense. It is interesting to see that this is close to optimal doping, that was discovered in experiments months later.

\begin{figure}[h]
\includegraphics[width=6cm]{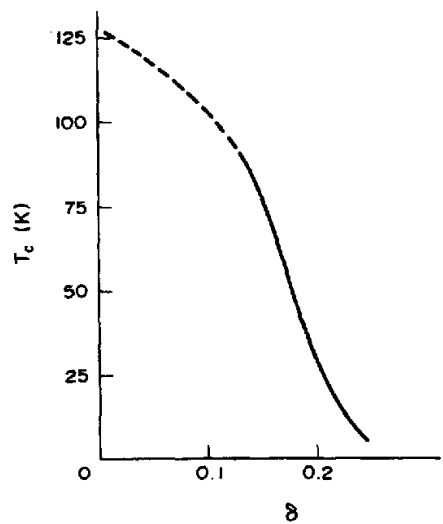}
\caption{Phase diagram reproduced from BZA\cite{BZA} paper. This phase diagram,drawn in March 1987, was based on a microscopic calculation plus some conjecture about validity of mean field theory at higher dopings. Surprisingly, the conjectured region of validity of meanfield theory (beyond dashed line) very nearly corresponds to optimal doping of $\sim$ 15 \% and beyond, that was experimentally confirmed later.}
\end{figure} 

At small doping superconductivity was viewed as a condensation of a fraction x of valence bond bosons that are charged and gave a BEC type of expression of \tc for small doping:   
\be
k_B T_c \sim \hbar^2\frac{2\pi}{m^*}\left(\frac{x}{2.61 v_0}\right)^{\frac{2}{3}}
\ee
Here m$^*$ is the effective mass of the charged valence bond and $v_0$ is the
volume occupied by each Cu cell. What is remarkable about this formula is that
the superconducting \tc has a strong and explicit dependence only on t and x. It does not have any explicit dependence on J or the Hubbard U !. The large U has done its job through superexchange, of preparing spin singlets at a sufficiently high temperatures. When BEC takes place below spin gap scale $k_BT_c^*$, spin singlet
correlation is maximal, and it is a coherent charge fluid, without much spin activity at low energies. This formula was modified, to a more appropriate Kosterlitz-Thouless type formula in 2D, in a subsequent paper\cite{pwa1} by with Anderson, Hsu and Zou,
as
\be
k_B T_c \sim \frac{2\pi \hbar^2}{m^*}(x - x_c)
\ee
Here x$_c$ is some critical doping needed to overcome disorder effects and begin
superconductivity; this expression was taken from a similar expression for superfluidity of $^4$He in vicor glass. At very high doping Mott insulator turns into a (disturbed) fermi sea. Superexchange becomes less relevant, as electrons are less localized because of decreasing correlations. This mean field theory showed a sharp decrease of the mean field \tc beyond an optimal doping. Synthesizing various ideas and the BZA mean field solutions, a phase diagram was suggested shortly, in the same paper\cite{pwa1}. This phase diagram, shown in figure 3, was also a prediction of BZA theory. The experimental phase diagram that was established later over years has a striking resemblance to this prediction.

\begin{figure}[h]
\includegraphics[width=8cm]{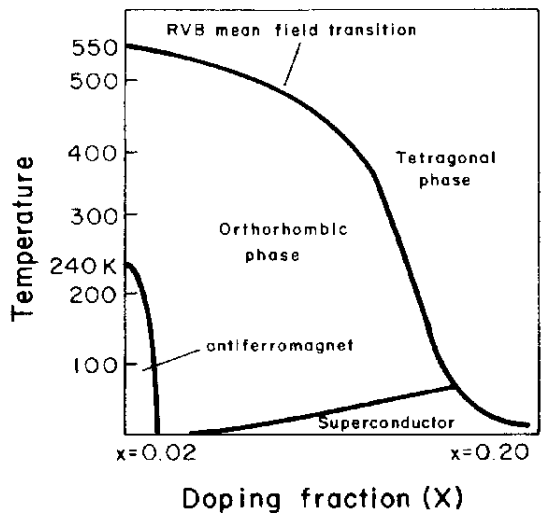}
\caption{Phase diagram reproduced from a paper by Anderson, GB, Hsu and Zou\cite{pwa1}. This phase diagram, drawn in April 1987, was based on a microscopic calculation plus some conjecture, before a systematic study of doping versus \tc began. The overall agreement with the current cuprate phase diagram is striking.}
\end{figure} 

This idea of bose condensation of charge valence bonds was given a sharper expression with important consequences, as a holon (soliton) condensation by Kivelson, Rokhsar and Sethna\cite{KRS} shortly. 

The importance of quantum phase fluctuations in a condensate of preformed pairs was emphasized in the BZA paper; these ideas/predictions which predated detailed experiments, have matured into notions such as phase fluctuation dominated bad metal phase\cite{emeryKivelson} and vortex liquid phase\cite{ongNernst}, with help from a variety of experiments performed after 1987.  This basic ideas from RVB
theory was used in an important paper by Emery and Kivelson\cite{emeryKivelson} to talk about quantitative comparision of superconducting \tc with experimentally measurable quantities, such as superfluid stiffness, using a phase fluctuation dominated scenerio. We will discuss this later.

It is also worth emphasizing that BZA paper constitutes an unequivocal prediction of the spin pseudo gap phenomenon, long before it was noticed experimentally.

Even though mean field theory was in agreement with experiments in giving high \tc superconductivity at optimal doping, an important question was whether the mean field superconductivity will survive quantum fluctuations arising from double occupancy constraint. this paper conjectured that when the doping is above 
$\approx 5 \%$ the fluctuation correction arising from the double occupancy constrant should not matter very much. However it was necessary to validate this in an acceptable fashion either using existing many body theory or some thing new.

\begin{flushleft}
\textbf{3. Gauge Theory}
\end{flushleft}

After BZA theory one of the urgent job was to give mathematical expression to
the increasing phase fluctuations of cooper pairs, as one approaches the Mott insulator by decreasing the doping. It also amounts to finding how to express mathematically difference between a Mott insulator and doped Mott insulator. The gauge theory paper\cite{gauge1} undertook this study from a very different perspective. It offered a new approach, which distinguished Mott insulator and doped Mott insulator and enabled a systematic study of Mott insulator based quantum fluctuations on superconductivity. This began a new activity of gauge theory approach to strongly correlated electrons systems, and in particularly cuprate superconductors. Some key ideas from lattice gauge theory was very effectively used to understand the strong correlation problems in Mott insulators and doped Mott insulators.

This theory observed that the low energy Hamiltonian of Mott insulator has a local
U(1) gauge symmetry, when expressed in terms of the underlying electron variables. This is manifest in the electron representation (equation 5).
A local (site dependent) gauge transformation
\be
c^\dagger_{i\sigma} \rightarrow
e^{\i \theta_i}c^\dagger_{i\sigma}
\ee
leaves the spin Hamiltonian invariant. Because, $b^{\dagger}_{ij}$ transforms as
\be 
b^{\dagger}_{ij} \rightarrow e^{\i \theta_i}b^{\dagger}_{ij}e^{\i \theta_j}
\ee
and leaves superexchange hamiltonian (equation 5) $ -J \sum_{\langle ij \rangle} b^{\dagger}_{ij} b^{}_{ij} $ invariant. An important consequence of this is a possibility of an emergent U(1) dynamical gauge field on the links. This link gauge field is a dynamically fluctuating complex field, $\Delta_{ij}(t)$, connecting 
neighboring sites ij. Around the same time, while studying Anderson lattice and heavy fermion problems, Noga\cite{noga} introduced such dynamically generated link variables. This insight was very useful for us and we used this effectively in our Mott insulator and doped Mott insulator problem. 

The dynamically generated field $\Delta_{ij}$ was shown in path integral treatment of the problem, using a Hubbard-Stratanovic method, which reduces a quartic two body interaction term to a quadratic interaction term with dynamically fluctuating pair field $\Delta_{ij}(t)$. Path integral expression for the partition function of the t-J model is:
\bearr
&~~~~Z& = \int  \prod_{i\sigma\tau}{\delta_{n_{i\uparrow}(\tau)n_{i\downarrow}(\tau),1}} dc^{\dagger}_{i\sigma}(\tau)dc^{}_{i\sigma}(\tau) \nonumber \\
&\times&e^{-\int^{\beta}_{0}d\tau \sum_{i\sigma}{c}^\dagger_{i\sigma}(\tau) (i\partial_{\tau} - \mu) c^{}_{i\sigma}(\tau)}
\nonumber\\
&\times&e^{-\int^{\beta}_{0} d\tau 
[- t\sum_{\langle ij\rangle} c^\dagger_{i\sigma}(\tau) c^{}_{j\sigma}(\tau) + h.c.
-J \sum_{\langle ij \rangle} b^{\dagger}_{ij}(\tau)b^{}_{ij}(\tau)]}
\eearr
Here $c^{\dagger}_{i\sigma}(\tau)$'s are Grossman variables; double occupancy constraint is formally expressed through a delta function along with the measure.
Using a Gaussian identity, $ e^{a^2} = \int^{\infty}_{-\infty} e^{-\pi x^2- 
2{\sqrt\pi}ax} dx$ we re-express the partition function as
\bearr
&Z& \sim  \int \prod_{\langle ij \rangle \tau} d\Delta^*_{ij}(\tau)d\Delta_{ij}(\tau) e^{-\int^{\beta}_{0} d\tau
J \sum_{\langle ij \rangle} |\Delta_{ij}(\tau)|^2} 
\nonumber  \\
&\times&\prod_{i\sigma\tau}\delta_{n_{i\uparrow}(\tau)n_{i\downarrow}(\tau),1} dc^{\dagger}_{i\sigma}(\tau)dc^{}_{i\sigma}(\tau)\nonumber \\
&\times&e^{-\int^{\beta}_{0}d\tau  \sum_{i\sigma}{c}^\dagger_{i\sigma}(\tau) (i\partial_{\tau} - \mu) c^{}_{i\sigma}(\tau)}
\nonumber\\
&\times& e^{-\int^{\beta}_{0} d\tau 
 [- t\sum_{\langle ij\rangle} c^\dagger_{i\sigma}(\tau) c^{}_{j\sigma}(\tau)  - J \sum_{\langle ij \rangle},b^{\dagger}_{ij}(\tau) \Delta_{ij} + h.c.]}
\eearr
The auxiliary variable $\Delta_{ij}(\tau)$ is the probability amplitude of finding a singlet bond between two electrons at sites ij. Now one can formally integrate over the fermi fields to get an effective action in terms of $\Delta_{ij}(\tau)$'s:

\be
Z \sim  \int \prod_{\langle ij \rangle \tau} D\Delta^*_{ij}(\tau)D\Delta_{ij}(\tau) e^{\int^{\beta}_{0} d\tau
S_{\rm eff}[\Delta_{ij}(\tau)]} 
\ee

Without performing an explicit integration one can make some very general and useful statement about the symmetry of the effective action $S_{\rm eff}[\Delta_{ij}(\tau)]$ at half filling and away from half filling.

At half filling the Heisenberg model exhibits a local U(1) gauge invariance. This implies that $S_{\rm eff}[\Delta_{ij}(\tau)]$ is invariant under the local U(1) gauge transformation:
\be
\Delta^*_{ij} \rightarrow e^{\i \theta_i} \Delta^*_{ij}e^{\i \theta_j}
\ee
We evaluated the static part of the effective action, taking care of the double occupancy constraint in a mean field fashion and obtained the following lattice action:
\bearr
S_{\rm Mott} &\sim& a_{0} \sum_{\langle ijkl\rangle} \Delta^*_{ij}\Delta_{jk}\Delta^*_{kl}\Delta_{li} + H.c. \nonumber \\
&+&b \sum_{\langle ij \rangle} |\Delta|^2 + c \sum_{\langle ij \rangle}
|\Delta_{ij}|^4
\eearr
This action exhibits a manifest local gauge invariance on the lattice. 
First term of the above action represents resonance of two valence bonds in an elementary square plaquette $\langle ijkl \rangle$. This is a large term, that is actually proportional to J, as was noted later by Anderson\cite{pwaPlaquette}. This term  
stabilizes resonating singlets in the ground state. When we focus on the soft
phase variable $\theta_{ij}$, we can approximate the hard variable, 
$|\Delta_{ij}| \approx \Delta_0|$ and  
$\Delta_{ij} \approx |\Delta_0| e^{i\theta_{ij}}$. In terms of the phase variable the U(1) action has the form:

\be
S_{\rm Mott} \approx c |\Delta_0|^4 \sum_{\langle ijkl \rangle}
cos (\theta_{ij}-\theta_{jk} + \theta_{kl} - \theta_{li})
\ee
This paper summarized results of an analysis, using Villain approximation, and a plaquette integer variable (similar to vorticity in XY model in 2D) was discussed. This integer plaquette variable was called `magnetic charges', using analogy to U(1) lattice gauge theory in 2+1 dimensions. These magnetic charges were identified as key `topological excitations' that will describe evolution of spin liquid state as a function of temperature. The magnetic charges or vorticity, is a novel topological excitation that arises in a disordered spin liquid state in 2D, the very first one as far as the authors knew. Kivelson-Rokhsar-Sethna's holon or spinon was shown to arise when $\Delta_{oj} = 0$, when they are located at the origin. 

It was further suggested that as the gauge field is dynamically generated, by
interacting fermions, the magnetic charges might induce fermions, through a possible topological term. This will be a `neutral fermion' (spinon) excitation.
In the modern language it will be called a fermion-flux composite. Very soon Dzhyalozhinksi, Wiegman and Polyokov\cite{polyakov} suggested Hopf term as a topological term. It suggested a possibility of statistics transmutation and a debate started about existence of Hopf term in 2D spin-\hlf Heisernberg antiferromagnet. In parallel, Marston\cite{marstonZ2} and also Zou\cite{zouChernSimon} suggested a Chern-Simon term in the U(1) gauge theory. Thus the gauge theory paper planted seed for the discussion of Chern-Simons field theory to describe 2D quantum spin systems.

Soon the magnetic charge was called by a more appropriate name, `magnetic flux' and
Affleck and Marson found the $\pi$-flux RVB mean field solution in the BZA theory. This state respects parity and time reversal, as $+\pi$ and $-\pi$ can not be distinguished quantum mechanically (Bohm-Aharanov effect). A chiral spin liquid state that was suggested by Kalmayer and Laughlin\cite{kalmayer}, in a triangular lattice antiferromagnet was shown to be related to an RVB mean field solution with condensed $\frac{\pi}{2}$ flux, by Feng and Lee\cite{fengLee}. In more recent works the magnetic flux was christened as `visons' in the works of Senthil and Fisher\cite{senthil}. Very recent work by GB\cite{gbSkyrmion}, following an early work of us with Anderson, John, Doucot and Liang\cite{johnMeron}, finds that the well known skyrmion solution of 2D Heisenberg antiferromanget represents two unbound spinons that carry quarter magnetic flux each, and showing an important result that spinons are deconfined semions, but above an energy gap.

The magnetic charge or flux, that came in a natural fashion in the RVB theory 
was described by Wen-Wilczek-Zee\cite{wwz} in terms of spin chirality variables expressed, in terms of Pauli spin operators as:
\be
 \Re ~~e^{i\oint\textbf{A}(\textbf{r}). d\textbf{l}} \sim \textbf{S}_i\times(\textbf{S}_j \times \textbf{S}_k)
\ee
Another important consequence of U(1) gauge field description in a lattice was an application of Elitzur's theorem\cite{elitzur}. Elitzur theorem states that in a pure lattice gauge theory (without matter) with local gauge invariance, the local gauge symmetry can not be spontaneously broken.
Thus Elitzur theorem automatically precludes spontaneous symmetry breaking of the U(1) gauge field in the Mott insulating state. It means absence of superconductivity in a Mott insulator. This paper also pointed out possibilities of confined and deconfined phases of U(1) gauge theory, through the area and power law behavior of Wilson loop like loop correlation functions 
such as $ W(C) = \langle \prod_C b^{\dagger}_{ij}b^{}_{jk}b^{\dagger}_{kl}
... b^{}_{pi} \rangle$. In the deconfined phase the neutral fermions carrying spin-\hlf moments were suggested to be present as low energy excitations. 

Next important question is the consequence of doping and how it will bring about superconductivity in the gauge field description. This paper showed that the doped Mott insulator described by the t-J model does not have local U(1) symmetry of the parent Mott insulator. That is, in the presence of a non zero doping x $\neq$ 0, the effective action
$S_{\rm eff}[\Delta_{ij}(\tau)]$ \textit{is invariant oly under a global U(1) gauge transformation}, $\Delta_{ij} \rightarrow e^{i2\theta} \Delta_{ij}$.

We evaluated the effective action, using the same approximation as before and obtained,
\bearr
S_{} &\sim& a_{0} \sum_{ijkl} \Delta^*_{ij}\Delta_{jk}\Delta^*_{kl}\Delta_{li} + H.c. \nonumber \\
&+& xa_1\sum_{\langle ijk} \Delta^*_{ij}\Delta_{jk} + H.c.\nonumber \\
&+&b \sum_{\langle ij \rangle} |\Delta|^2 + c \sum_{\langle ij \rangle}
|\Delta_{ij}|^4
\eearr
This had a remarkable consequence. As the second term proportional to x, the dopant density, removes the local gauge invariance, total action has only a global U(1) gauge symmetry. The first term representing a plaquette resonance of the valence bonds is the memory of the Mott insulator in this approach. An immediate consequence was that Elitzur's theorem is no more applicable now. In principle superconductivity is possible. Action (equation 31) is the simplest lattice Ginzburg Landau action for RVB superconductors. The major aspect of RVB appears from the plaquette resonance term, which fights against long range order. During 1988, Nakamura and Matsui\cite{matsui} used the above lattice action and did a complete numerical evaluation of the partition function (going beyond saddle point approximation) and found reasonable superconducting Tc close to doping. This was an important numerical
proof that 2D superconductivity survives gauge field fluctuations at optimal dopings.

Soon after the gauge theory paper, Muller-Hartman's group\cite{mullerHartman} calculated the coefficient of the static lattice RVB-GL action, taking care of double occupancy constraint more accurately and obtained a detailed phase diagram in the x-T plane.

The effective action found the gauge theory paper, like the BZA method, has the right physics as far as symmetry of the superconducting order parameter is concerned. The coefficient of the `gradient term' in the lattice 
$ xa_1\sum_{\langle ijk} \Delta^*_{ij}\Delta_{jk} + H.c.$ had the right sign 
( xa$_1 > 0$). And minimization of the above auction automatically leads to a d$_{x^2-y^2}$ solution, very similar to Kotliar's d-wave solution. 

The RVB-GL action above does not have fermions explicitly, as they have been integrated out completely. This is not correct for a d$_{x^2-y^2}$ superconductor as it has nodal fermion quasi particles. So the low energy Bogoliubov quasi particles and their coupling to $\Delta_{ij}$ should be present as part of the action. This is easy to incorporate either phenomenologically or microscopically. Many authors such as, Affleck, Marston, Matsui, Nakamura, Wiegman, Ioffe, Larkin, Patrick Lee, Wen, Wilczek, Zee, Nagaosa, Read, Sachdev, Balents, Nayak, Fisher, Senthil, Dung-Hai Lee, Tesanovic, Franz and others have contributed to the elaboration of these fundamental ideas, most of them using continuum action.   

However, it is important to point out that it is difficult to incorporate the plaquette term, which keeps the memory of the Mott insulator, in a continuum approximation. To this extent, the RVB-GL theory on a lattice remains unexplored. We find that because of  the plaquette term, which distinguishes it from a standard superconductor, many interesting consequences could occur; for example, Andreev bound states at the vortex core and nature of impurity states induced by Zn and Ni substitution at copper site. In a recent work Muthukumar and Weng\cite{muthuGL} have developed an RVB-GL theory, starting from a slave fermion approach, and studied the physics of spinon vortices and properties of electromagnetic vortex core.

\begin{center}
\textbf{\large Completing the BZA Programme}
\end{center}

\begin{flushleft}
\textbf{1. d-wave Solution in BZA theory}
\end{flushleft}

As mentioned earlier, the available experimental data at the beginning of 1987 was not inconsistent with absence of long range magnetic order in the Mott insulating end. Further, in the doped Mott insulator, experiments continued to show a small
amount of electronic linear specific heat at low temperatures in the  superconducting region. Both these results gave PWA a confidence that the quantum spin liquid with a pseudo fermi surface at the Mott insulator continues to become a superconductor with extended-s symmetry, and retains neutral fermionic excitations with a pseudo fermi surface. Overwhelmed by this confidence we were sailing happily in a psuedo fermi sea.  

During this period, intrigued by the structure of BZA mean field theory and the gauge theory\cite{gauge1} approach, Affleck and Marston\cite{piFlux} found a $\pi$ RVB flux mean field solution for the Mott insulator and Kotliar\cite{kotliar} the d$_{x^2-y^2}$ solution for the superconductor, within the BZA approach. Affleck and Marston presented their result as an exact result in a large N limit of a generalized Heisenberg model in 2D. 

As mentioned earlier, mean field energy of d-wave being lower than extended s-wave is seen as follows. In BZA theory, the pair potential, for pair scattering, is given by $- J\gamma({\bf k - k'}) \equiv - J[cos(k_x-k_x') + cos(k_y-k_y')]$. This potential, which is attractive for small momentum transfer ${\bf k-k'}$, changes sign and becomes repulsive for large momentum transfer ${\bf k-k'} \approx (\pm \pi,\pm \pi)$. The d$_{x^2-y^2}$ solution, which changes sign as we move along the fermi surface in k-space takes advantage of this and becomes a lower energy state.

Even after Kotliar's solution, there was a reluctance from Princeton group to accept d-wave solution. In addition to the then existing phenomenological support for extended-s solution, we were not sure if the BZA d-wave solution will continue to have lower energy than extended-s solution, after Gutzwiller projection.
Later Gutzwiller projected d-wave BCS solution was shown to have lower energy
using numerical methods. 

The beauty of Affleck-Marson and Kotliar's solution was that both had nodal quasi particles. Various experimental results that followed later demostrated clearly d-wave superconductivity for hole doped cuprates. Needless to say that an unconventional order parameter such as d-wave has its own profound consequences for cuprates, as has been seen both in the experimental and theoretical fronts in the last two decades. It is fair to say that RVB theory contains in its bag all these fascinating possibilities and perhaps even more. A recent excitement is with respect to superconductivity in Na$_x$CoO$_2 \cdot $yH$_2$O. We has developed an RVB theory for this system\cite{gbCOB}, where d + id, another unconventional order parameter that violates parity and time reversal symmetry is predicted. 

\begin{flushleft}
\textbf{2. Improving BZA mean field Hamiltonian}
\end{flushleft}

As discussed earlier, the hard problem of Gutzwiller projection was replaced by
an ansatz in the variational analysis. We replaced the hopping parameter t by 
tx. This tells us that an electron can hop to a neighboring site only when it is empty. The probability of it being empty is x, the doping density. This is equivalent to replacing bare electron mass by a renormalized mass, which was already familiar to us from the work of Brinkman and Rice in the context of Mott insulator to metal transition in Hubbard model at half filling. 

While evaluating expectation values involving Gutzwiller projection, an approximate method that takes care of some incoherent aspects of projection, was developed by
Gutzwiller, in his earlier study of Hubbard model. It involved certain combinatorics. This method was already successfully used by Rice,Joynt, Shiba, Ogatta, Volhardt, Anderson and others in dealing with issues of heavy fermions and also Hubbard modeling of solid $^3$He. This method was adapted by Zhang, Gros, Rice and Shiba\cite{ZGRS} to improve the BZA renormalization process. Their improvement is schematically shown below. 
\be
P_G (H_t + H_s) P_G~~~~ \stackrel{\rm BZA}\rightarrow ~~~ (x~H_t + H_s) ~~~
\stackrel{\rm ZGRS}\rightarrow ~~~(g_t~H_t + g_s~H_s)
\ee
with doping dependent renormalization parameters $g_t = \frac{2x}{1+x}$ and 
$g_x = \frac{4}{(1+x)^2}$. But for the two new renormalized parameters the rest of the self consistent theory was identical to BZA theory. Further, physical correlation functions such as anomalous amplitudes $\langle b^{\dagger}_{ij}\rangle$, got an appropriate renormalization factor g$_t$ times mean field amplitude :
\be
\langle b^{\dagger}_{ij}\rangle \rightarrow g_t\langle b^{\dagger}_{ij}\rangle_0
\ee
This is in spirit similar to BZA theory, where it is stated that fraction of singlet bonds that are charged are x and use this to calculate superconducting 
\tc. It should be also pointed out that the renormalization parameters 
g$_t$ and g$_s$ are also not variational parameters, very much like in BZA theory.
Here also one gets a gap equation for the order parameter $\Delta$ 
and p, which can be solved self consistently. The merit of the improvement suggested by ZGRS is that the results obtained by a mean field anaylsis is quantitatively more accurate. This is elaborated in the Plain Vanila paper
as well as more recent works of Zhang et al and Muthukumar-Gros et al. 

Generalization of the above BZA program to finite temperature has problems. This problem is related to variational states that are orthogonal before Gutzwiller projection becoming non-orthogonal after Gutzwiller projection and decrease in density of relevant Hilbert space introduced by double occupancy constraint.
Anderson\cite{pwaSpinLock} has recently offered an analysis, where he introduces the notion of `spin-charge locking' and introduces a generalized BCS type of formalism, where separate Anderson-Nambu spinors are introduced to take care of neutral 
spin-pairing and electron pairing. At very high temperatures they are decoupled and in the superconducting state they are locked. Spin-gap phase represents a progressive locking. 

\begin{flushleft}
\textbf{3. Holon Condensation \& Slave Boson theory - \\
a support for BZA program}
\end{flushleft}

Soon after Anderson's paper and BZA theory, an insightful paper by Kivelson, Rokhsar and Sethna\cite{KRS} offered a theory where a doped charge enters as holon,
a spin zero soliton with charge +e. They act like bosons and bose condense
leading to superconductivity. It seemed to express, in a formally correct fashion, Anderson and BZA's result of viewing the charged valence bond as a boson and their bose condensation. Zou and  Anderson\cite{za} adapted the slave particle method, developed earlier by Barnes, Coleman, Read, Newns, Kotliar and Ruckenstein, to the t-J model. This began a series of investigations starting with the work of Kotliar-Liu,
gauge theory by Wiegman\cite{wiegman} and the present author\cite{gauge2}, and detailed mean field analysis by Fukuyama\cite{fukuyama} and collaborators and several other investigations. 

The advantage of slave particle method over the variational approach is that one can introduce dynamically generated gauge fields and go beyond mean field theories, without constrained by variational RVB states and also address finite temperature problems. However they are technically hard, as it is clear in the works of many authors that followed. BZA mean field theory seems to capture the essence of superconductivity phenomena in the t-J model.

In some limit the slave boson analysis gives the same result as the improved BZA Hamiltonian by Zhang et al. During this development, we also showed\cite{pwa1} 
that the holon condensation is not actually a charge 2e rather charge e condensation in view of the double occupancy constraint. In other words, electron pairs get effectively delocalized into a zero momentum condensed state. Or holon is book keeping device for a correlated fluctuation and delocalization of charged valence bonds, as far as superconductivity is concerned.

\begin{flushleft}
\textbf{4. Lurking dangers outside Optimal Doping}
\end{flushleft}

As RVB theory was being developed, there was a conscious effort to focus on region around optimal doping. RVB superconductivity is at its best and one can hope to understand it better here, than elsewhere. In fact, \textbf{t-J model is a reasonable model for real cuprates, when we have either an isolated hole in a Mott insulator or a finite density of holes at optimal doping !} Any thing in between is complicated because of unscreened long range coulomb interactions among charge careers, strong coupling to phonons and disorder effects from off plane dopants. It could easily lead to nano scale phase separation into Mott insulating and optimally doped regions, self trapping etc. 

In reality it turned out to be even more complicated. A doped Mott insulator also supports other quasi or psuedo long range orders such as fluctuating charge and spin orders and chiral orders, involving circulating spin currents and charge currents. Interactions not contained in the t-J model, seem to encourage these competing orders, outside the superconducting dome. They are fascinating quantum and classical condensed matter problems. However, they are not crucial in our foremost goal of understanding high \tc superconductivity deeply. There are good phenomenological and theoretical evidences that they are not the root cause of high \tc superconductivity. In fact, they are competitors\cite{gbStripe}

The same is true of the so called pseudo gap phase, outside the dome. Experimental evidences point to presence of enhanced correlations corresponding to the cometing order. 

There is a clear indication that the physics and interactions contained in the t-J model 2D is alone able to support high \tc superconductivity, without need for an external help or a catalytic agent.

\begin{center}
\textbf{\large Mean field theory works remarkably well at optimal doping}
\end{center}

RVB mean field theory and gauge theory approach, predicted a robust superconducting state at optimal doping with a large \tc and supported, Anderson's proposal. 
These papers also pointed out that in the superconducting region, mean delocalization energy per particle ($\sim xt$) and energy of superexchange (J) are comparable. The superconducting condensation energy is a finite fraction of J or $xt$. Now we know that $\frac{J}{k_B} \sim 1500 K$ for cuprates and superconducting condensation energy is large and so are superconducting \tc's.  This puts cuprates on a different region in the Uemura plot. In fact, Uemura plot itself was inspired by RVB type of idea of bose condensation of charged valence bond in cuprates. 

Fortunately for the above theories, new cuprate family members were discovered, where superconducting \tc's soared to new heights. The Hg and Tl based single layer superconductors reached a Tc in the range of 95 K. At optimal doping, \tc is large compared to its one layer counterpart \lsco or one layer BISCO material. I have argued elsewhere, that a single layer material has a large intrinsic superconducting \tc $\approx 120 K$. Competing orders (charge and spin order tendencies, often helped by octahedral rotation or distortions) steal away the superconducting condensation energy, making superconducting \tc among one layer materials to swing from 5 K to 95 K at optimal doping. 

The highest Tc is in the Tl based cuprates, where Tc is as large as 163 K, under large external pressures. These are remarkable experimental support to the BZA
and gauge theory approximations. 

Several theoretical attempts readily show enhanced singlet correlations in the ground state, a prerequisite for long-range singlet superconductivity. Variational Monte carlo analysis of RVB superconducting wave functions have given useful results, where quantitative comparisons have been made with some experimental quantities. Over years quantum Motne carlo methods have given encouraging results.
Other approaches, such as real space as well as k-space cluster DMFT have given
very meaningful results, consistent with BZA RVB mean field theory. 

One of the best support for BZA program was provided by a semi phenomenological theory due to Emery and Kivelson\cite{emeryKivelson}, which focused on superconducting transition temperature. Kivelson and Emery assumed, consistent with RVB theory and the existing phenomenology that superconducting phenomena at low doping is dominated by phase fluctuations rather than amplitude fluctuations. This leads to a simple expression for superconducting \tc in terms of measurable quantities. In particular
energy associated with spatial variation of phase is expressed as:
\be
H \approx  \frac{\hbar^2n_s(0)}{8m^*}\int(\nabla \theta(\textbf{r}))^2 d\textbf{r}
\ee  
Here m$^*$ is the effective mass of cooper pair and n$_s$(0) is the superfluid density at zero temperatures. Then one can use the standard KT formula to get an expression for superconducting
\tc as
\be
k_BT_c \approx \frac{\pi \hbar^2 n_s(0)}{8 m^*}
\ee
This expression is remarkably similar to the expression given in BZA paper
and the later paper with Zou and Hsu (equation 20 and 21). Emery and Kivelson
went further and made detailed comparison with existing phenomenology, including
how \tc increases with inter layer coupling etc. The remarkable result is that,
as in the BZA result, here also large U or J does not explicitly appear in the
final result for \tc. Superexchange, a consequence of large U, simply provides
a large energy scale and an umbrella below which fairly stable singlets can delocalize and produce a superconducting state, along the lines of BEC.
As thermally produced nodal quasi particles will interfere with superconductivity,
in a way different from the fully gaped superconductors, certain corrections were
necessary to the above analysis, as indicated in reference\cite{pleeReview}.

Even though no rigorous theorem exists proving existence of a finite \tc superconductivity in large U repulsive Hubbard model or t-J model in 2D, it is increasingly becoming clear that such a theorem is likely to exist. However, one 
should remember that spin-\half systems are notoriously hard. For example, a rigorous demonstration of long range antiferromagnetic order, in the undoped 2D square lattice Heisenberg antiferromagnet does not exist even now. Will doping make it any simpler ?

As mentioned earlier, in the heat of developments during the beginning of 1987, a phase diagram was conjectured in the paper by Anderson, Hsu, Zou and the present author, based on BZA theory (figure 3). This phase diagram was conjectured well before experimental phase diagram emerged. It is interesting that the experimental phase diagram had an excellent overal form as conjectured by the theory. This again indicates that the fluctuation effects do not change the qualitative prediction of the mean field theory. 

BZA paper also pointed out that ` ... in the low doping concentration limit (x $<<$ 1) phase fluctuations play an important role and the mean field theory fails.
At this limit, \tc and gap are governed by the large phase fluctuations. ...
But for large enough x ($> 5 \%$) we expect that the mean field theory works.'
The main reason for the conjecture was that, in order to be able to construct a phase coherent superconducting state, two most important requirements are i) well developed short range singlet correlations and ii) finite long wavelength charge compressibility. Singlet correlation is there in plenty in the doped Mott insulator, because super exchange continues to be present for a range of doping. A doping of 5 $\%$ gives sufficient delocalization energy to escape localization effects due to disorder or hole-hole electrostatic interaction and attain a finite charge compressibility. 

Spin-1 collective mode at $(\pi,\pi)$ first seen in neutron scattering\cite{keimer},
is one of the key unconventional feature of cuprate superconductivity. As far as I know no such spin-1 collective mode is seen in any conventional superconductor. Moreover, the frequency of this mode scales linearly with superconducting \tc. In fact, when \tc becomes zero this mode becomes the Goldstone mode of the antiferromagnetic order, that is supported by a Mott insulator through superexchange. Thus spin-1 collective mode is a \textbf{memory of the Mott insulator}. It is a simple manifestation of the tightly bound bond singlets that makeup the superconducting state: these singlets are softer at $(\pi,\pi)$ in k-space. This reveals a dynamic antiferromagnetic correlation at $(\pi,\pi)$, at energy scales large compared to the superconducting energy scale. The simplest theory that explains the spin-1 collective mode at $(\pi,\pi)$, in a natural fashion at $(\pi,\pi)$ is  an RPA collective mode analysis built on a BZA type of d-wave mean field background, within a t-J model. This is another support for the mean field theory.

In a recent paper Anderson and collaborators\cite{vanila} have made comparison of the physical properties calculated within ZGRS approach, variational Monte-Carlo analysis and experimental results. The agreement between the three
are very good, underscoring the validity of the original BZA approach and the 
assumptions therein.

Another reason why a simple BCS type of theory works well at optimal doping
is the following. For zero and very small doping the strongly correlated electronic system has strongly localized electrons and there is no resemblance of the ground state to a metallic fermi sea. For very high doping we clearly have a degenerate 
fermi gas; at these heavy dopings superexchange does not take place, as electrons have a large mean velocity and do not have time for superexchange interaction with a neighbor in real space. At optimal dopings a reasonable fermi sea is formed at the same time superexchange also survives. It is this combination which makes the situation very similar to a BCS theory with superexchange being the glue. This is the reason BZA renormalized Hamiltonian and the improved version
by ZGRS work well.

\begin{center}
\textbf{Is Spin Fluctuation a Glue ?}
\end{center}

The BZA program that was initiated in 1987 and got completed shortly, is capable of answering many questions in the superconducting region. Surprisingly such work began only recently. Partly because, we were either ignorant or ignored key old developments. Too many questions, not necessarily related to the major debate, arose from a wealth of experimental results. Being a complex system, cause and effects get mixed up sometimes. We will use the spin fluctuation scenario to 
illustrate our point.

Historically, Hirsch's numerical analysis of the single bond Hubbard model,
brought out the beautiful possibility of an extended-s wave superconducting correlation. This and other developemens started focusing on this phenomena in terms of diagrammatics and exchange of spin fluctuation bubbles.

Let us look at experimental facts. At optimal doping, a spin-1 resonance
emerges at ($\pi,\pi$), as a sharp mode, but only in the superconducting state.
In the normal state it becomes very broad and disappears. This is a collective mode, characteristic of the RVB superconducting state. Further, the energy of
this mode scales linearly with superconducting \tc. Formally, as the superconducting \tc vanishes, this mode becomes a spin wave or Goldstone mode.
Further, in addition to the spin-1 resonance there is a broad background of 
spin fluctuation activities, as seen in S(q,$\omega$). 

RVB theory takes the point of view that spin flucuation contribution to S(q,$\omega$) around ($\pi,\pi$) is a result of a built up of singlet 
correlation through superexchang processes. These fluctuations that occur in a broad
range in momentum and energy space are far from any coherent modes. Instead of talking about these dissipative spectrum of spin fluctuation activities, RVB theory directly focuses on superexchange J and treats it as a glue. The only coherent mode is the spin-1 resonance. As it has been pointed out, it has a very small
spectral weight and further it occurs only below \tc. So it can not be a glue
either.

All these phenomena are manifest in diagrammatic spin fluctuation calculations. Electron has a strong frequency and momentum dependent and large normal and anomalous self energies. They depend on each other self consistently. The large normal part of the self energy makes the quasi particles spectral function very broad, consistent with ARPES experiments. But this approach misses the built up of singlet correlations, as a result there is no way we can even approach under doped region, leave alone the Mott insulator region by this approach. There is no natural and simple way of getting superfluid
density being proportional to doping at small x. 

There is a missing logic. While one may get some satisfactory answers, a complete picture, a claimed strength of RVB theory, is absent.

\begin{center}
\textbf{\large Conclusion}
\end{center}

In condensed matter physics we attempt to synthesize new laws, notions and introduce new reference states or phases in demystifying properties of complex materials. There are several idealized reference states for describing a variety of quantum phenomena in solids: free electron gas, ideal bose gas, harmonic phonons, fermi liquids, Luttinger liquids, BCS paired fermi sea, etc. The wealth of quantum condensed matter phenomena force us to introduce new reference states occasionally. 

PWA's RVB proposal in 1987 and subsequent developments in the last 20 years illustrates a struggle to introduce Mott insulator as a reference phase to describe the unusual high Tc superconductivity and a variety of related anomalies\cite{pwabook} in the family of cuprates. From superconductivity point of view, it is a serious attempt to find an alternative to phonon mediated superconductivity, compelled by experiments.

Looking back, it has been a worthwhile struggle, and RVB theory has silently 
entered the subconscious of the condensed matter mind. \lco has become a text book Mott insulator. A one band model, with strong correlation, namely t-J or large U Hubbard model is accepted as a minimal model to describe low energy physics of cuprates. Deep consequences of projection in these models are also getting accepted, but slowly. The U(1) RVB gauge field theory\cite{gauge1} has been nurtured and developed by several authors in commendable ways. Old ideas from RVB are being rediscovered. The pseudo fermi surface and neutral fermions suggested  PWA\cite{pwascience} and shown to be a possibility in a microscopic theory in BZA, is being realized in certain organic Mott insulator, ET-salts\cite{kanoda}. RVB mechanism is being successfully applied to understand superconductivity in organics\cite{gbOSC,zhangEtAl}. A new superconductor Na$_x$CoO$_2\cdot$ yH$_2$O, is likely a long awaited doped spin-\hlf Mott insulator on a triangular lattice\cite{gbCOB}. There is a strong indication for RVB superconductivity in boron doped diamond\cite{gbDiamond}, through superexchange effects in an impurity band Mott insulator. 

In the field of superconductivity, before cuprates appeared in the scene, one was used to the luxury of a beautiful and powerful BCS theory, that is so useful in understanding majority of elemental superconductors. Formalisms such as the Eliashberg theory, a microscopic approach and other related developments have been extremely helpful in understanding many experimentally measured properties, including tunneling spectra, $\alpha^2F(\omega)$ etc., accurately. Ginzburg-Landau phenomenological theory gets a microscopic meaning through Nambu-Gorkov formalism. Space and time dependent superconductivity phenomena and quasi particle dynamics is understood through Bogoliubov-de Gennes theories. There are new phenomena that came as prediction after BCS theory: Josephson effects, Andreev reflection etc. All is well with old superconductivity. Once we agree to live with a few parameters of less microscopic origin, even some bad actors such as A15 and Chevral phase compounds seem to yield to BCS theory.

The situation with cuprates and many new materials, suspected to be RVB superconductors are different. We mentioned about the lurking dangers out side optimal doping from competing phases and extra interactions that could encourage their growth. This makes a theory based on t-J model of some what limited validity !
Much care should be taken to get all the low energy physics from the t-J model. 
For example, determination of superconducting \tc for a given system, say
\lsco is wrought with complications in the way we explain below. 

In BCS theory we used to worry about isotope shift of a fraction of a degree in \tc. In cuprates, at optimal doping, in one layer materials, the \tc varies between 5 K to 95 K between Bismuth and Thallium one layer materials. All these materials have very similar normal state properties, such as the coefficient of linear resistivity, modulo some material purity complications. Such a large variation indicates that superconductivity is not alone. There are other factors and competitors that are at work. Single layer superconducting \tc may even get enhanced by an interlayer pair tunnelling phenomena\cite{wha} in bilayer systems, an entirely new additional contribution that owes its origin to anomalous normal state.

So the world of cuprate superconductors are different and complex. First one has to answer some generic questions as accurately as possible for the 2 dimensional t-J model. For example, what is the x dependence \tc near optimal doping. What is the maximum \tc at optimal doping for a range of t and t' and J that is relevant for cuprates. How \tc gets modified in a bilayer or multilayer system for a given $t_{\perp}$. 

It is not meaningful to pick up \lsco and try to understand superconducting \tc from
t-J model.  We will be tempted to find the correct band parameters t, t', J and do 
an accurate RVB mean field theory. It is very clear that what controls the experimentally seen reduced \tc is some phenomena outside t-J modeling or band structure effects. In fact, in one experiment\cite{naito} an epitaxial strain increases the superconducting \tc of \lsco thin film from 25 to 50 K, without any manifest increase in doping. Simple estimate shows that epitaxial strain can not give sufficient change in t's to cause such a large effect. The fundamental reason seems to be that epitaxial strain is encouraging certain atomic scale lattice distortion that encourages charge order formation self consistently.

There is some new physics\cite{gbStripe}, namely a competing phase, which is stealing away superconducting condensation energy. As BISCO and Tl one layer materials are affected in different ways, actual theoretical prediction of ground state gap parameter, will require additional inputs, either phenomenological or microscopic. One of the important question is how easily the Cu-Oxygen octahedron, or pyramid or square planar complex respond and get distorted or rotate in their different environments, to either charge or spin or valence bond localizations and ordering tendencies. Then there will be feed back and growth of competing orders,
at the expense of superconductivity.

One of the strategies will be to pick the best superconductor in the single layer family for a deeper understanding of mechanism of superconductivity. From this point of view our first superconductor namely doped \lco is a bad system to study ! Superconducting \tc never goes beyond 30's at optimal doping. Do we have to abandon two decades of experimental efforts ? Perhaps not. Then an important question is why doped \lco, or other low \tc members have never attained their full potential (maximum \tc) that Tl one layer system has reached. These are new questions that has no parallel in conventional superconductors.

At the end it may be possible to accommodate effects of competing orders phenomenologically by modifying t', J etc. But that might miss important physics. It is important to recognize and treat competing dynamical processes in their own right. This of course makes the problem hard.  

So we realize after 20 years of cuprate study that we are in a different situation.
The nature of questions asked should be different and method of analysis will be
different. Priorities will be different. For example one of the important question will be how to reach the full potential or maximum superconducting \tc in the one layer family ? As we said earlier, the one layer superconducting \tc can be as large as 95 K. However, in multi layers we have a superconducting \tc of 163 K. Is it a consequence of cooperation from interlayer pair tunneling phenomena within the multi layers, or  some structural rigidity that discourages competing orders, there by making a single layer realize new heights ? 

We have focused on only one thermodynamic property. We have similar question about the energy of the spin-1 collective mode and a variety of other key physical properties.

There are several important experimental issues in the superconducting state: temperature dependence of order parameter, nodal quasi particle dynamics, as revealed by magnetic resonance studies, S(q,$\omega$) from neutron scattering, spectral functions of quasi particles in ARPES, detailed space and energy dependent STM study of local electronic density of states and gap functions, quasi particle interference effects, structure of vortex core, states inside the vortex core, bound states around impurities such as Zn, Ni and their different and unexpected effects as a function of temperature, thermal conductivity anomalies etc. etc. 

Suddenly one finds oneself in the midst of a flood of questions and real complications, so different from elemental superconductors. As we mentioned earlier it is a meeting place of quantum magnetism and superconductivity, Mott insulator and fermi sea. Many things other than superconductivity are taking place. It is unfair to say we do not understand high \tc superconductivity. We understand it too well, so we fear it and tread carefully. 

Anomalous normal state phenomena takes us to a different world all together. 
Experimentally, it is a clear case for a non fermi liquid in 2D. The many body theory we have developed, to address these challenges, are at the beginning stage. PWA has attacked this problem from different angles : i) a non-vanishing  phase shift at 2k$_F$, in the forward scattering spin singlet channel, in 2D Hubbard model ii) tomographic Luttinger liquid, asymptotic Bethe ansatz in 2D iii) Anderson-Khveschekno's anomalous commutation relations, iv) spin charge decoupling and two relaxation times on the fermi surface v) asymmetry in single particle tunneling arising from the key projection in a t-J model and vi) very recently an orthoganality catastrophe inherent in the projected t-J model etc.  
 
Similarly, in defect free underdoped cuprates superconductivity seems to vanish
because of quantum fluctuations or unscreened coulomb interactions, leaving a metallic ground state. This possibility that was suggested earlier\cite{BZA} with a pseudo fermi surface and later with nodal quasi particles\cite{fisherNodalQP}. The one with nodal excitations seems to be gaining experimental support\cite{talifer} in very pure YBCO. It will be a pristine RVB state, a reference non fermi liquid state, that has not yielded to instabilities such as charge order or spin order or superconductivity. P and T violating metallic ground states are likely to exist in organics and in cobalt oxide systems\cite{gbCOB}, near the Mott insulator end.

So it is humbling to see the kind of problems and also richness that Bednorz and M\"{u}llers discovery and Anderson's RVB proposal has created in the world of superconductivity and strongly correlated electronic systems. At the same time it is heartening to see that BZA and related papers have answered the primary question of existence of high temperature superconductivity in cuprates in a microscopic theory, rather satisfactorily, and is ready to answer several old and new questions.

\begin{center}
\textbf{Acknowledgement}
\end{center}
I thank P.W. Anderson for a continuing collaboration and for insightful and critical discussions. I thank M. Muthukumar (Amherst) and V. N. Muthukumar (CUNY) for very useful comments on the manuscript.

\end{document}